\documentclass[conference]{IEEEtran}
\IEEEoverridecommandlockouts
\usepackage[utf8]{inputenc}
\usepackage{graphicx}
\usepackage{float} 
\usepackage{tikz}
\usepackage{xcolor}
\usepackage{caption}
\usepackage{subcaption}
\usepackage{hyperref}
\usepackage{placeins}
\usepackage{refcount}
\usepackage{comment}
\usepackage{booktabs} 
\usepackage{multirow}


\usepackage{amsfonts}
\usepackage{pifont}
\newcommand{\cmark}{\ding{51}}
\newcommand{\xmark}{\ding{55}}
\newcommand{\ocircle}{\ding{109}}%

\definecolor{lime}{HTML}{A6CE39}
\DeclareRobustCommand{\orcidicon}{
	\begin{tikzpicture}
		\draw[lime, fill=lime] (0,0)
		circle[radius=0.16]
		node[white]{{\fontfamily{qag}\selectfont \tiny \textbf {\.{I}D}}};
	\end{tikzpicture}
	\hspace{-2mm}
}
\foreach \x in {A, ..., Z}{%
	\expandafter\xdef\csname orcid\x\endcsname{\noexpand\href{https://orcid.org/\csname orcidauthor\x\endcsname}{\noexpand\orcidicon}}
}

\definecolor{dodgerblue}{RGB}{64, 67, 194}
\newcommand{\Reviewed}[1]{\textcolor{black}{#1}}
\title{Lifecycle Management of Trustworthy AI Models in 6G Networks: The REASON Approach
}
	\author{%
		\IEEEauthorblockN{
			\parbox{\linewidth}{\centering
				  Juan Parra-Ullauri\IEEEauthorrefmark{1},
                    Xueqing Zhou\IEEEauthorrefmark{1},
                    Shadi Moazzeni\IEEEauthorrefmark{1},
                    Rasheed Hussain\IEEEauthorrefmark{1},
                    Xenofon Vasilakos\IEEEauthorrefmark{1},
				Yulei Wu\IEEEauthorrefmark{1},
                    Renjith Baby\IEEEauthorrefmark{2},
                    M M Hassan Mahmud\IEEEauthorrefmark{2},
                    Gabriele Incorvaia\IEEEauthorrefmark{3},
                    Darryl Hond\IEEEauthorrefmark{3},
                    Hamid Asgari\IEEEauthorrefmark{3},\\
                    Andrea Tassi\IEEEauthorrefmark{4},
                    Daniel Warren\IEEEauthorrefmark{4},
				 and Dimitra Simeonidou\IEEEauthorrefmark{1} %
			}%
		}%
	\thanks{\hspace{-0.3cm}\IEEEauthorrefmark{1} High-Performance Networks Group, Smart Internet Lab University of Bristol, BS8 1QU, UK (e-mail: jm.parraullauri@bristol.ac.uk),\protect\\\IEEEauthorrefmark{2} Digital Catapult, AI and Data Science Technology,  NW1 2RA, London, UK (e-mail: renjith.baby@digicatapult.org.uk),\protect\\\IEEEauthorrefmark{3} Thales UK, Research Technology and Solution Innovation, Reading, UK (e-mail: darryl.hond@uk.thalesgroup.com),\protect\\  \IEEEauthorrefmark{4} Communication Solutions at Samsung R\&D Institute UK (SRUK), UK \protect\\(e-mail: a.tassi@samsung.com).}
	}

\begin{document}

\maketitle

\begin{abstract}
Artificial Intelligence (AI) is expected to play a key role in 6G networks including optimising system management, operation, and evolution. This requires systematic lifecycle management of AI models, ensuring their impact on services and stakeholders is continuously monitored. While current 6G initiatives introduce AI, they often fall short in addressing end-to-end intelligence and crucial aspects like trust, transparency, privacy, and verifiability. Trustworthy AI is vital, especially for critical infrastructures like 6G. This paper introduces the REASON approach for holistically addressing AI's native integration and trustworthiness in future 6G networks. The approach comprises AI Orchestration (AIO) for model lifecycle management, Cognition (COG) for performance evaluation and explanation, and AI Monitoring (AIM) for tracking and feedback. Digital Twin (DT) technology is leveraged to facilitate real-time monitoring and scenario testing, which are essential for AIO, COG, and AIM. We demonstrate this approach through an AI-enabled xAPP use case, leveraging a DT platform to validate, explain, and deploy trustworthy AI models.
\end{abstract}

\section{Introduction}
\label{intro}
As 5G deployment accelerates, 6G research aims to address future demands driven by digitisation and hyper-connectivity~\cite{hexaXVision}. 6G networks will be distributed, intelligent, and closer to end-users, supporting global coverage, security, and efficiency. Unlike 5G, which supports only radio access technology, 6G will enable multi-access technologies (mATs)~\cite{reason}, requiring native intelligence embedded in the network fabric. AI will play a key role in optimising networks, improving resource allocation, fault detection, mAT control, and maintenance, while enhancing security and reducing costs~\cite{hexaXVision}. However, integrating AI into 6G poses challenges, particularly in managing AI lifecycles and ensuring trustworthiness in critical infrastructures~\cite{reason}.
Trustworthy AI involves creating AI systems that are reliable and accountable~\cite{huang2020survey}. It encompasses principles and technologies that ensure AI functions align with human values, respect privacy and fairness, and maintain transparency and explainability. Recognising its importance, various organisations globally are setting requirements for AI systems, with the EU’s AI Act being a prominent example. Trustworthy AI is essential for future intelligent 6G networks, ensuring that AI systems used in network operations and decision-making are dependable and avoid severe human, financial, or legal consequences.

To fill the identified gaps, this paper presents the REASON~\footnote{\Reviewed{For more information about the REASON project and the Architectural proposal visit: \url{https://reason-open-networks.ac.uk/}}} project's approach towards native and trustworthy intelligence, focusing on training, deploying, and updating AI models, ensuring adherence to high standards of integrity and accountability in the future open network (including 6G)~\cite{reason}. The approach comprises three main components: AI Orchestration (AIO), Cognition (COG), and AI Monitoring (AIM). AIO manages the AI model lifecycle, including cataloguing versions, descriptions, and requirements. It automates model training, manages data, and supports deployment and scaling across environments. COG assesses AI model performance holistically, verifies AI system behaviour and ensures alignment with expectations while communicates anticipated performance and explanations to stakeholders targeting Trustworthy AI. Finally, AIM involves probes distributed across the network to collect data on running AI models to build a Digital Twin (DT), which is then fed back to the AIO and COG components. These components ensure AI models improve and stay relevant by analysing their performance and impact on systems and stakeholders as network conditions evolve.

We will discuss the importance of AI in 6G networks, outline the fundamental principles of trustworthy AI, and compare REASON's approach with existing 6G projects in Section~\ref{preliminaries}. In Section~\ref{proposal}, we will introduce the REASON architecture and provide a detailed explanation of the roles and interactions of AIO, COG, and AIM. Additionally, we will demonstrate the feasibility of the approach in Section~\ref{example} by considering the development, validation, and implementation of an xAPP for application content classification based on user-generated traffic from mATs such as WiFi, LiFi, and cellular. The challenges associated with the seamless integration of AI and trustworthiness will be discussed in Section~\ref{challenges}. Finally, we will conclude this article in Section~\ref{conclusion}.

\section{Background and Related Work}
\label{preliminaries}

\subsection{6G Networks and AI}
6G is set to revolutionise communication systems, enabling scenarios such as immersive communication, massive and ubiquitous connectivity, hyper-reliability, low latency, AI integration, and integrated sensing and communication as outlined in the ITU IMT-2030 Framework. Key features of 6G include speeds up to 100 times faster than 5G, microsecond-level latency, and support for massive connectivity with densities of $10^6$-$10^8$. Advanced applications like Metaverse, robotic surgery, truly immersive environments and teleportation will also be prominent. The increased complexity and demands of 6G necessitate incorporating AI models for effective network optimisation, enhanced security, efficient resource allocation, and personalised user experiences. Traditional ``patch" AI solutions add complexity due to the lack of systematic AI model management. Therefore, integrating native AI into 6G is essential for lifecycle management, automatically monitoring AI's impact on networks, services, and stakeholders, and enabling necessary adaptations.

\subsection{Trustworthy AI}
In essence, a trustworthy entity or system will behave as intended and according to general principles and interests. Trustworthy design requirements include robustness, reliability, resilience, explainability, safety, security, privacy, and fairness. In technical terms, a component, subsystem, or system can be designated trustworthy if the above requirements have been met and verified. For an AI-enabled system to be deemed trustworthy, it must meet the given criteria for trustworthiness at the component, subsystem, or system level~\cite{huang2020survey}. Trustworthy AI delivers transparent AI-based decision-making, bolsters user acceptance, stakeholder trust, and confidence in AI capabilities, and ensures compliance with laws and regulations governing AI usage and ethical frameworks, thus supporting a responsible deployment of AI applications. 

AI-native 6G networks must be trustworthy and will function autonomously to address diverse social needs. Thus, trustworthy AI is crucial for the successful implementation and operation of these networks. It guarantees that the networks are robust, transparent, secure, fair, and adhere to ethical and legal standards. Prioritising trustworthiness allows the development of networks that fulfil technical requirements while earning the trust and confidence of users and stakeholders, thus promoting the adoption of AI-native autonomous networks.

\subsection{Related Work}
In recent years, several initiatives have proposed integrating AI into networks. The REASON project introduces a novel architecture aligned with the 6G landscape discussed earlier, advancing the goal of embedding native AI in future networks with a focus on trustworthiness~\cite{reason}. Table~\ref{tab:comparison} highlights key contemporary contributions and positions REASON relative to existing initiatives across various dimensions and categories.

\begin{table*}
    \renewcommand{\arraystretch}{1.3}
    \caption{Comparison of Contemporary Projects that Consider (Trustworthy) AI and Networks}
    \centering
\begin{tabular}{l c c c c c c c}  
    \hline 
    \multicolumn{1}{c}{}  & D1: ML Lifecycle Mgmt & D2: MLOps & D3: ML Resource Prediction & D4: DTs for NI & D5: Monitoring & D6: TAI Framework \\
    \hline \hline 
        HEXA-X   & \cmark & \cmark & \ocircle  & \cmark & \cmark & \ocircle\\
        ADROIT6G  & \cmark & \xmark & \xmark & \xmark & \cmark & \ocircle  \\
        DAEMON   & \cmark & \cmark & \xmark  & \xmark & \cmark & \ocircle   \\
        DESIRE6G & \cmark & \cmark & \xmark & \xmark & \cmark & \ocircle   \\
        RIGOROUS   & \cmark & \xmark & \xmark &  \cmark & \cmark & \xmark \\
        ACROSS  & \cmark & \xmark & \xmark &  \cmark & \cmark & \ocircle  \\
        NANCY  & \xmark & \xmark & \xmark &  \xmark & \cmark & \ocircle  \\
        PREDICT6G & \cmark & \cmark & \ocircle &  \cmark & \cmark & \xmark   \\ 
        6G-BRICKS  & \ocircle & \cmark & \xmark &  \xmark & \cmark & \ocircle \\ 
        \textbf{REASON}   & \cmark & \cmark & \cmark & \cmark & \cmark & \cmark   \\
        \hline
\end{tabular}
\vspace{1ex}
\label{tab:comparison}

\parbox{0.8\textwidth}{
    \centering \cmark\ Feature Present \quad \xmark\ Feature Absent \quad \ocircle Partially Addressed
}

\end{table*}

The native intelligence depends on \textbf{D1: Machine Learning (ML) Lifecycle Management}, which focuses on developing, deploying, and removing ML models within the network, covering data collection, training, and inference. Various projects in the literature have recognised this as a crucial aspect when integrating AI into the network, as illustrated in Table~\ref{tab:comparison}. To manage the lifecycle of ML models, projects such as HEXA-X~\cite{hexaXVision}, DAEMON~\cite{daemon6g}, and PREDICT6G\footnote{\label{fn:predict6g}PREDICT6G - D3.1 Release 1 of AI-driven Inter-domain Network Control, Management, and Orchestration Innovations.} utilise \textbf{D2: ML Operations (MLOps)}, which is a structured framework of practices and principles designed to streamline and automate the entire ML lifecycle. REASON considers these dimensions and emphasises model selection, model chaining, prioritisation, and \textbf{D3: ML Resource Prediction} within the context of AIO which has not been considered in the literature. \Reviewed{ML resource prediction is a process for forecasting the resources required for running ML tasks, particularly in heterogeneous environments like 6G networks. In REASON, we address this through AI Model profiling~\cite{parraullauri2024profilingaimodelsefficient} further described in Section~\ref{proposal_AIO}.}

The \textbf{D4: DT for Network Intelligence (NI)} category features diverse approaches. RIGOROUS focuses on network security testing using DTs\footnote{\label{fn:rigorous}RIGOUROUS - D4.1: Design Plan of the AI-Driven Anomaly Detection, Decision and Mitigation.} whereas PREDICT6G predicts network analytics with DTs\footnotemark[\getrefnumber{fn:predict6g}]. REASON aims to develop a cutting-edge DT for a 6G network, integrating mATs, facilitating offline AI model training, and incorporating lifecycle management, explainability, and verifiability modules for heterogeneous scenarios. In the \textbf{D5: Monitoring} category, HEXA-X offers a comprehensive monitoring framework for seamless software integration and data exchange~\cite{hexaXVision}. At the same time, RIGOROUS proposes a real-time data processing framework with an emphasis on privacy\footnotemark[\getrefnumber{fn:rigorous}]. REASON will also address energy consumption, distributed monitoring, and data integrity~\cite{reason, matching_service}.

Various projects demonstrate different cognitive capabilities and approaches related to \textbf{D6: Trustworthy AI (TAI) Framework}. HEXA-X prioritises Explainable AI (XAI) frameworks for predictive tasks, focusing on transparency and reliability in AI decision-making~\cite{hexaXVision}. On the other hand, ADROIT6G balances performance and explainability in network slicing scenarios~\cite{androit6GExplainAI}. \Reviewed{The 6G-BRICKS project promotes XAI in 6G networks by enabling transparent, AI-driven resource orchestration and automated network management with clear decision insights.} Another similar project NANCY aims to improve the explainability of self-supervised deep clustering models for wireless spectrum activity~\cite{milosheski2023xai}. On the other hand, REASON prioritises explainability, robustness, and privacy in AI from a holistic standpoint. This framework ensures correctness, reliability, and adherence to ethical standards across all its AI/ML operations. Additionally, REASON incorporates comprehensive AIM capabilities within its architecture embedding Trustworthy AI principles from the architecture and design perspective. 

In summary, the REASON approach enhances the AI plane concept by embedding Trustworthy AI principles including verifiability and explainability within the ML pipeline, focusing on certified models that account for system and stakeholder impacts. The framework introduces distinctive elements, including AI model profiling, ML resource prediction, and lifecycle management tailored for heterogeneous 6G environments. REASON's approach towards native and trustworthy AI in future networks will be presented in the following section.

\section{Trustworthy AI for 6G Networks: \\ The REASON Approach}
\label{proposal}
\subsection{REASON Approach towards AI-Native 6G}

\begin{figure*}
    \centering
    \includegraphics[width=0.85\textwidth]{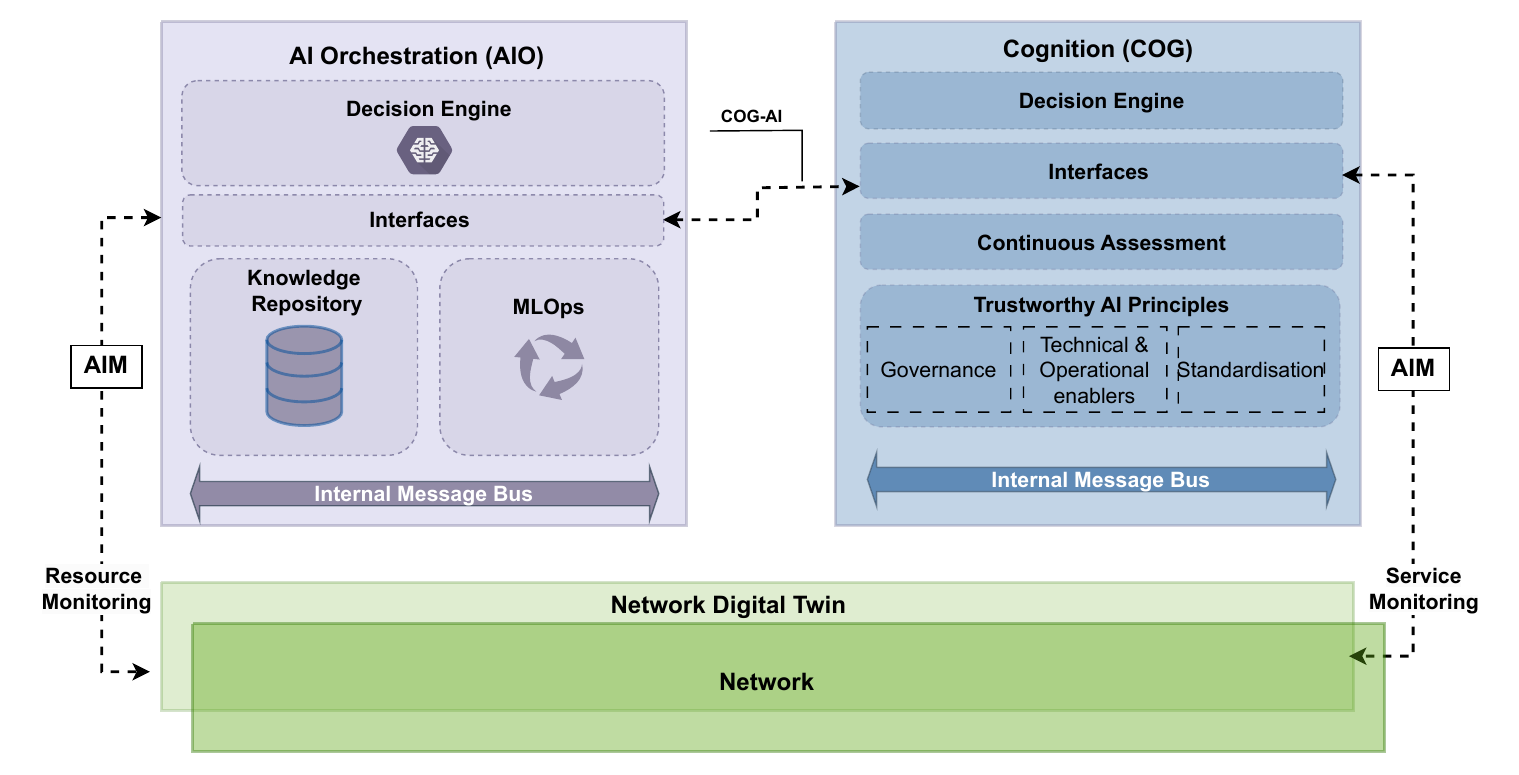}
    \caption{End-to-end AI Plane: AIO, COG, AIM and DT}
    \label{fig:aio_cog_dt}
\end{figure*}

AI-native 6G networks require a comprehensive, end-to-end system for managing the full lifecycle of AI models, from development to retirement. This lifecycle involves defining the problem, acquiring and processing data, model training, deployment, monitoring, retraining, and replacement. Achieving native intelligence demands a holistic system- and service-level perspective to optimise data collection, model distribution, and deployment across the network. AI-native 6G networks must also assess model performance in service delivery, enable model reuse for varied applications, and verify and explain model behaviour to ensure reliability, transparency, and adaptability within evolving network environments. \\
In REASON, our main objective is to create an AI-Native network architecture by providing AI/ML models across the network and ensuring that the performance of AI models is in line with the expectations~\cite{reason}. The E2E AI plane (shown in Figure~\ref{fig:aio_cog_dt}) comprises three main components AIO, AIM, and COG. AIO refers to the automated management, coordination, and optimisation of the AI model lifecycle, including versioning, training, deployment, monitoring, and retirement. AIM refers to probes distributed across the network to collect data about running AI models. COG provides a holistic analysis of AI model performance and impact within a system focusing on ensuring Trustworthy AI. These components will be detailed in the following sections, with a particular emphasis on COG to ensure trustworthy AI in 6G.

\subsection{Lifecycle Management of Trustworthy AI Models}
\subsubsection{\textbf{AIO}}
\label{proposal_AIO}
REASON's AIO in 6G networks involves lifecycle management and control of AI models via an AI Orchestrator. This component maintains a catalogue of registered AI models, handling version control, automated training pipelines, data management, deployment, and scaling across various environments. It ensures model performance monitoring, anomaly alerts, and robust security and access controls. The Orchestrator also manages model deprecation and retirement, providing scalability and high availability. \Reviewed{AIO incorporates MLOps practices to streamline and automate the end-to-end ML lifecycle, ensuring seamless coordination and trustworthy deployment of AI solutions. Key tasks include selecting the right AI models with the help of COG, placing models in conjunction with Network Orchestrator, chaining AI models, allocating computational resources (CPUs, GPUs, TPUs), and utilising split learning and distributed computing.} AIO is related to D1, D2 and D3 in Table~\ref{tab:comparison}. 

\Reviewed{Our initial work on AIO focuses on profiling AI models across diverse network fabrics to assess the requirements for running AI models on different network nodes. AI profiling examines how device dynamics and system characteristics impact model performance across varied hardware and environments~\cite{parraullauri2024profilingaimodelsefficient}. This analysis aims to reveal relationships among model types, hyperparameters, hardware specifications, and dataset characteristics, and their effects on accuracy, resource utilisation, and task completion times. Profiling these elements enables accurate prediction of system behaviour under different configurations and workloads, which is essential for optimising resource use and maintaining high service quality.}

\subsubsection{\textbf{AIM}}
\Reviewed{For AI-native 6G networks, effective monitoring of AI-driven systems is essential to optimise performance, enhance security, and enable efficient resource management. Unlike using AI to aid network monitoring, AIM focuses on monitoring the AI models themselves. Strategically deployed probes continuously collect data on critical performance metrics such as latency, throughput, accuracy, and resource usage. This real-time data offers insights into AI operations, facilitating informed decision-making and proactive performance adjustments. Additionally, the gathered data aids in optimising AI model deployment (i.e., AIO) and improving AI understanding (i.e., COG), helping ensure ethical compliance and regulatory alignment in complex 6G environments.}

\Reviewed{In the REASON project, we propose a distributed, cloud-native, and service-based monitoring approach. This approach leverages microservices, each responsible for specific functions, with upper-level orchestration coordinating microservice connectivity. Our initial efforts concentrate on monitoring the network fabric where AI models operate, specifically assessing the resources these models require for optimal operation and performance~\cite{parraullauri2024profilingaimodelsefficient}. This granular, distributed monitoring solution enables flexible scaling, efficient resource allocation, and improved performance oversight across network nodes~\cite{matching_service}.} AIM is related to D4 and D5 in Table~\ref{tab:comparison}.

\subsubsection{\textbf{COG}}
In REASON, COG refers to functions that analyse intelligent applications, including AI, in 6G networks. These functions may involve human intelligence, automatic processing, or human-machine teaming. Key COG functions include 1) reasoning about AI applications for 6G orchestration, considering computational cost, reliability, and performance; 2) establishing the trustworthiness of AI components by verifying privacy, robustness, performance, explainability, and compliance; 3) receiving and evaluating explanations from AI components; 4) providing strategies for conflict resolution among simultaneous AI demands, and 5) interacting with AIO. COG is related to D4 and D6 in Table~\ref{tab:comparison}. The three pillars of REASON's trustworthy AI approach are privacy preservation, explainability, and verifiability \emph{by-design} and will be detailed next.

\paragraph{Privacy-preservation}
REASON explores Privacy Enhancing Technologies (PETs) to help organisations meet ‘data protection by design’ obligations. PETs are particularly suitable in contexts that involve large-scale collection and analysis of personal data. For example, AI-enhanced networks, IoT, and Cloud Computing Services. To this end, we investigated the following PETs:
\begin{itemize}
    \item Data Minimisation PETs: These reduce or remove individual identifiability, weakening the link between original and derived data. Examples are Differential Privacy (DP) and synthetic data.
    \item Data Protection PETs: These shield data while preserving its utility. Examples include homomorphic encryption and zero-knowledge proofs, which support security principles.
    \item Access Control PETs: These manage access to confidential data, ensuring minimal personal data sharing while maintaining data confidentiality and integrity. Examples include trusted execution environments, secure multi-party computation, and federated learning and analytics (FL, FA).
\end{itemize}

REASON's initial work on PETs focused on FL/FA and DP, developing a framework for deploying FL/FA in cloud-native networks with Isolation-by-design and Differentially Private Data Consumption~\cite{parra2024kubeflower}. 

\paragraph{Explainability}
XAI provides justifications for its outputs in terms that human users easily understand. To enhance trustworthiness, an AI application must generate accurate and clear explanations. This capability ensures that users can better understand and trust the AI's decisions, which is crucial for applications in complex systems such as 6G networks. In addition to improving user trust, explanations play a vital role in verifying AI outputs, enabling users to have greater confidence in the system's adherence to specific requirements. Explainability also involves uncertainty quantification, providing users with insights into the reliability of the AI's outputs and helping them make informed decisions. 

XAI is typically divided into three categories: inherently interpretable models, such as linear regression and decision trees, which offer straightforward explanations; model-agnostic techniques, like LIME and SHAP, which can be applied to various models for flexible explanations; and model-specific methods that are tailored to particular types of models, such as ensemble-based approaches and deep learning models. In the use case section, we will demonstrate how model-agnostic explanation modules can be embedded in the lifecycle of an AI model. 

\paragraph{Formally Verifiable AI}
Formal Verification of AI (FVAI) models offer a precise and deterministic method for ensuring AI trustworthiness \cite{huang2020survey}. This approach adapts tools and techniques from formal software verification to AI systems, aiming to create mathematical proofs that AI models meet specific requirements defined using formal languages like first-order logic. FVAI tools verify whether these properties hold true, addressing well-known instability issues in modern deep neural networks \cite{szegedy:2013}. Such instabilities can significantly impact critical infrastructures, such as telecom networks, making FVAI crucial for validating AI models used in 6G networks.

While FVAI provides deterministic guarantees, its current capabilities are limited to verifying properties related to output variations given input changes. This framework is particularly effective for checking model robustness, as it can handle any DNN behaviour expressible in these terms. For example, we demonstrate the use of FVAI in our experiments (Section~\ref{example}) to verify properties of content classification based on network traffic across mATs.

\paragraph{Verification by Testing}
Confidence in an ML model's compliance with requirements is often achieved through testing rather than formal verification. Unlike formal methods, testing does not provide proof but offers empirical and statistical evidence of performance. Typically, a trained ML model is evaluated on unseen test data similar to the training data to assess generalisation. Requirements specify the model's performance level on this dataset, and testing verifies compliance. Additionally, model trustworthiness includes robustness to perturbed data. Perturbed test data can be generated through parameterised transforms or synthetic data techniques~\cite{hendrycks2019benchmarking}. A robustness requirement specifies performance criteria for such data, which can be verified through testing.

A systematic approach must be taken to compiling test datasets taking into account the following factors (some of which are interrelated): dataset bias, coverage of the input space and domain representativeness; data sampling methods and sampling densities; and dataset class balance and size. Once a dataset has been generated, there are further criteria that can be used to assess its adequacy for testing a trained ML model, such as the degree to which it exercises the internal elements of the model - namely the test coverage. For example, test coverage metrics have been formulated to record the extent to which artificial neural network neurons are activated by a test dataset~\cite{pei2017deepxplore}.

\subsection{The Role of DTs in Native AI and Trustworthiness}

As 6G aims to integrate advanced technologies and services, ensuring reliability, security, and service quality is critical. DTs, virtual replicas of physical entities or systems, offer a dynamic environment to analyse network behaviours and AI-driven functionalities in real-time~\cite{reason}. Enhancing AI models' trustworthiness is vital for DT systems in 6G networks. In REASON's approach, DTs are essential for AI model lifecycle management, from development and validation to deployment and optimisation. By creating a digital counterpart of the physical network, operators can monitor AI model performance under various scenarios, ensuring compliance with trustworthiness criteria and enhancing AI resilience and robustness.

DTs enable a comprehensive approach to trustworthiness by providing a controlled environment to stress-test AI models and evaluate responses to varying network conditions, cyber threats, and operational challenges. Observing AI responses helps identify its strengths and weaknesses. This proactive strategy helps maintain 6G AI-native network reliability and stability. Additionally, the continuous feedback loop between the physical network and its DT allows real-time updates and improvements, ensuring AI models adapt to evolving network dynamics and threats. Figure~\ref{fig:aio_cog_dt} shows the interaction between AIO, COG, and network DTs, highlighting the importance of DTs in maintaining network integrity without risking actual network components.

\section{Example scenario: Trustworthy AI-enabled xAPP instantiation}
\label{example}
In this section, we demonstrate REASON's approach for the Lifecycle Management of Trustworthy AI Models in 6G networks through a use case of mAT networks. \Reviewed{Support for mATs sits at the heart of 6G (and beyond) networks and therefore, intelligent control and optimisation is essential beyond the existing Intelligent Controller (IC) in the current 5G networks. To address this, REASON envisions a new multi-Access Technology Real-Time Intelligent Controller (mATRIC)~\cite{reason}. mATRIC incorporates new microservices (in addition to the traditional xApps in O-RAN compliant architecture) to control and optimise mATs. Without loss of generality, the main objective of the current work is the instantiation of AI-enabled xAPP using a DT platform to verify, explain, and deploy trustworthy AI models.}
\begin{figure*}
    \centering
    \includegraphics[width=0.85\linewidth]{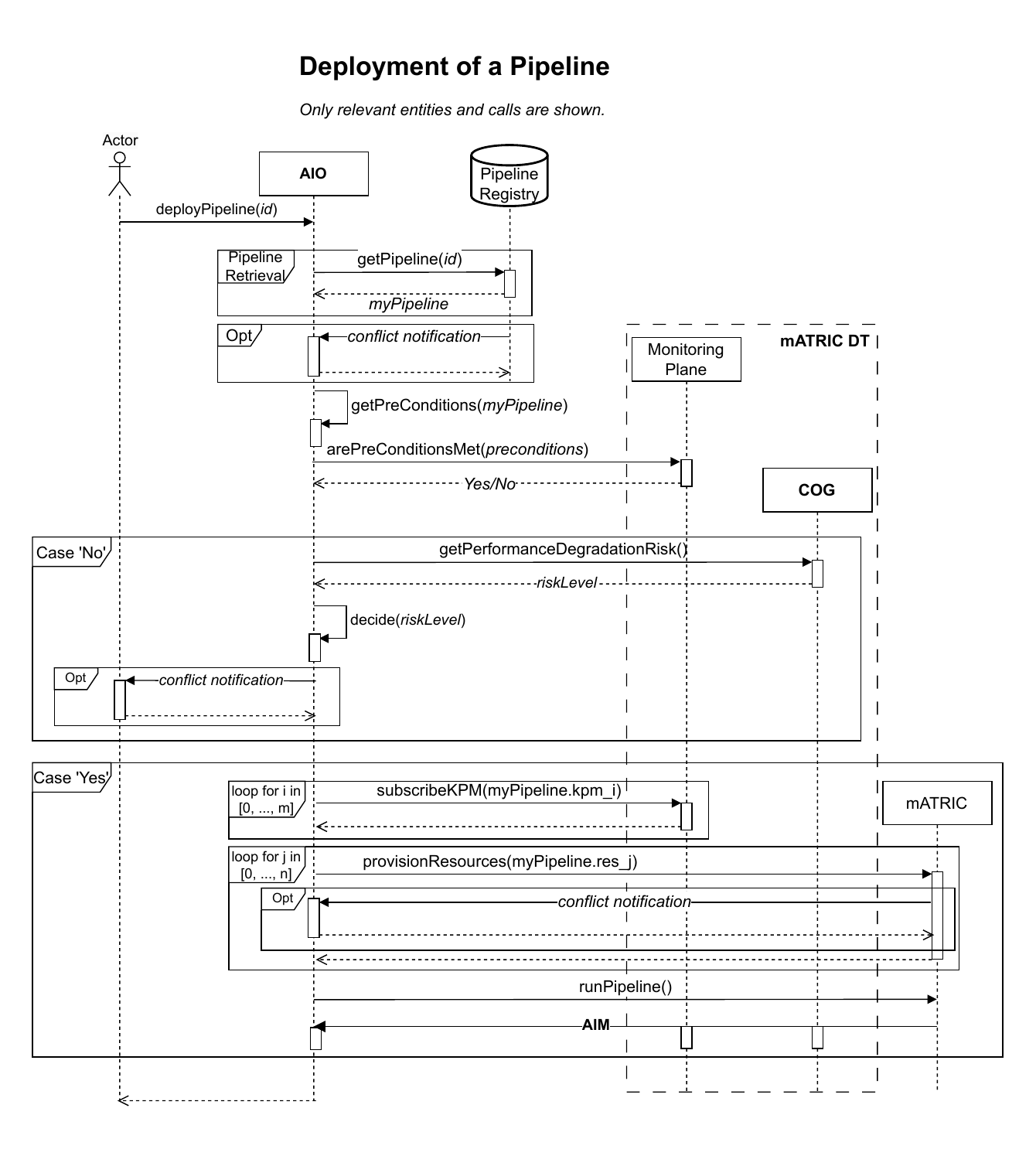}
    \caption{Example Deployment of an AI Pipeline using AIO, COG, AIM and the mATRIC DT}
    \vspace{-0.2cm}
    \label{fig:DT sequence diagram}
\end{figure*}

\subsection{Platform: Digital Twin of mATRIC}

To demonstrate the practical use of AIO, COG, AIM, and DTs in enhancing trustworthiness within 6G networks, we present a case study of our DT for mATRIC. This DT leverages data from the mAT in our testbed or simulation models to develop functional xApps for network control and optimisation. It includes four main components: database replication, synchronised network modelling, infrastructure and MLOps workflow orchestration, and ML-specific monitoring. The mATRIC platform monitors the mATs and stores data in its database, with synchronisation achieved via InfluxDB replication. The MLOps pipeline, using Kubeflow, manages data collection, preprocessing, model training, and trustworthy analysis. Network modelling is handled by the NS3 simulator, which tracks testbed configurations and updates the simulation module. Monitoring tools like MLflow provide insights into training performance and trustworthy analysis.

\subsubsection{\Reviewed{Deployment of an AI Pipeline}}

\Reviewed{Figure~\ref{fig:DT sequence diagram} illustrates the deployment of an ML pipeline using a sequence diagram involving AIO, COG, AIM, and the mATRIC DT. When a pipeline request is initiated, the AIO issues a \texttt{deployPipeline()} call with a specified pipeline ID, beginning with a check for pipeline availability via \texttt{getPipeline(id)} from the Pipeline Registry. If unavailable, a conflict notification is sent to the actor. When available, the AIO retrieves the pipeline's deployment requirements through \texttt{getPreConditions(id)}, detailing the necessary resources like CPU, GPU, and memory. To ensure a successful deployment, AIO then interacts with the Monitoring Plane, verifying resource readiness through \texttt{arePreConditionsMet()}. If not met, COG at the DT evaluates the associated risks and advises the actor.\\
If the required conditions are met, AIO subscribes to Key Performance Monitoring (KPM) metrics via \texttt{subscribeKPM()} and provisions the necessary resources through \texttt{provisionResources()}. The mATRIC DT proceeds to deploy the pipeline, preparing essential components like containers to support execution. Once deployment is complete, AIM probes are activated to monitor performance, and the pipeline is initiated with a \texttt{runPipeline()} command, launching the model and its operations. This structured approach, combining monitoring, precondition verification, and risk evaluation, ensures that the pipeline deployment is optimised for performance and aligned with available system resources.}

\begin{figure*}[!t]
    \centering
\includegraphics[width=0.95\linewidth]{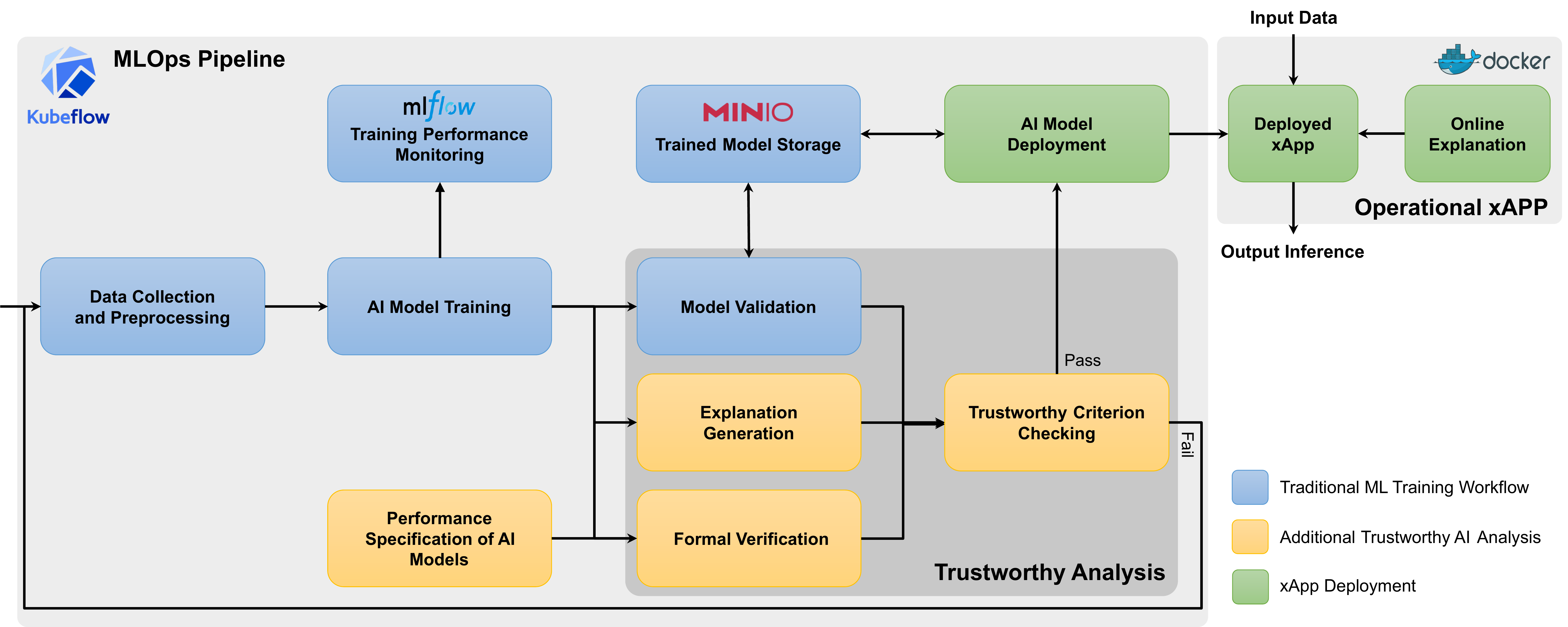} 
    \caption{Testbed Deployment of an MLOps Pipeline with Trustworthy Evaluation}
    \label{fig:ml-pipeline}
\end{figure*}

\subsubsection{\Reviewed{Deployment of an MLOps Pipeline in DT}}
\Reviewed{The MLOps pipeline, which oversees the AI model lifecycle, is depicted in Figure~\ref{fig:ml-pipeline}. Deployed by Kubeflow, this pipeline incorporates trustworthy AI analysis modules, such as formal verification and explanation components. Initially, the necessary data is retrieved from the database using its index. The gathered raw data is preprocessed and then trained using a specified MLP algorithm. The training performance and trustworthy performance is monitored using MLflow. The resulting model is saved in a MinIO repository, which generates a URI link to the model. After successfully passing the trustworthy criteria checks, the model is deployed as a Docker container in mATRIC as an xAPP for online inference-making alongside an online explanation module. REASON's AI approach prioritises explainability and verifiability as detailed in Section~\ref{proposal}. }

\subsection{Experiment Definition and Evaluation}
\subsubsection{Dataset, Model, and Task}

\begin{figure}[!t]
    \centering
    \includegraphics[width=0.85\linewidth]{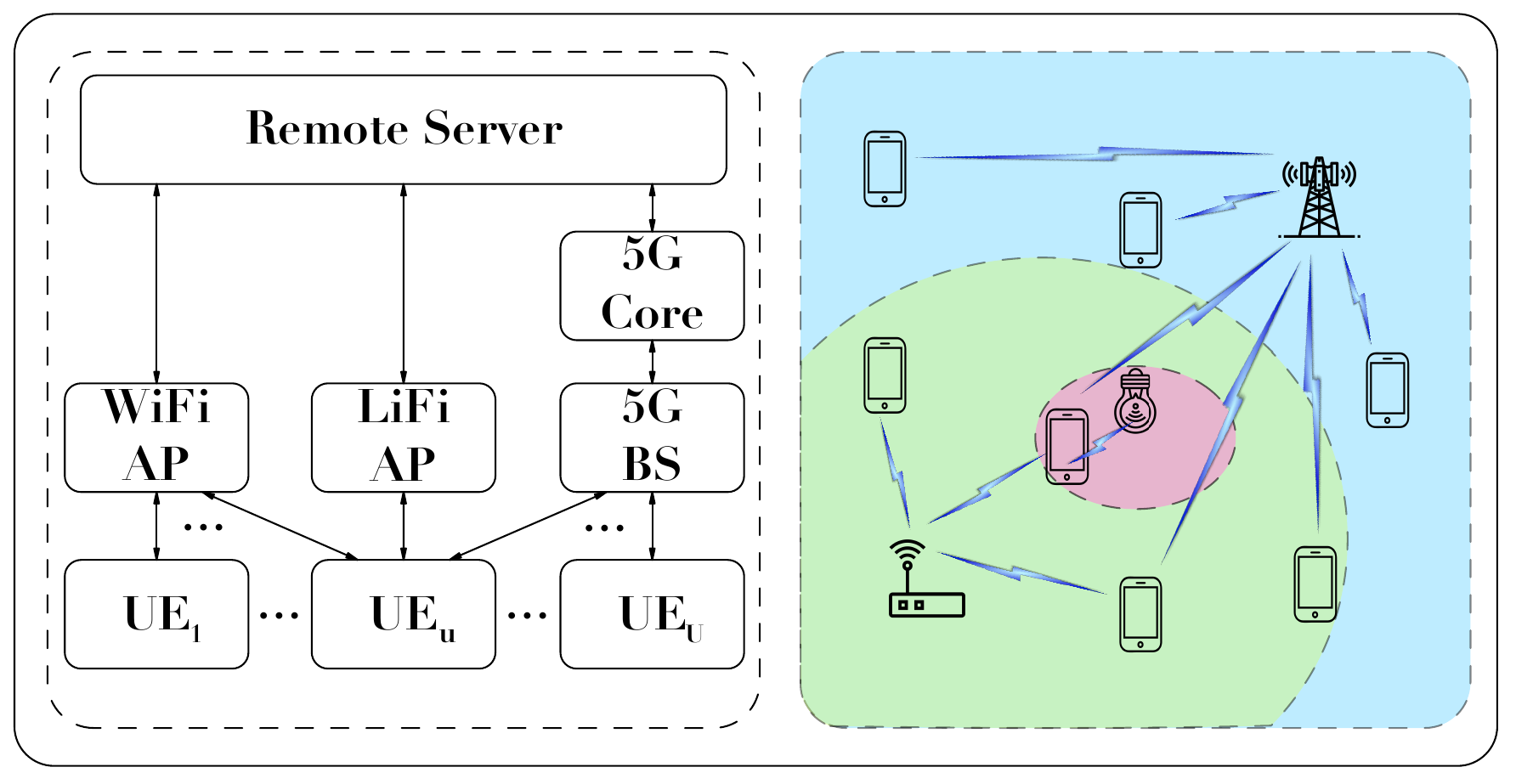}
    \caption{mAT Experiment in NS3 Simulator}
    \label{fig:ns3-sim}
\end{figure}

In this paper, we utilise the simulation module in the DT of mATRIC to showcase the lifecycle management of trustworthy AI models. The simulation scenario, as depicted in Figure~\ref{fig:ns3-sim}, emulates future network environments with varying numbers of UEs distributed in a 70-square-metre area. Access technologies include a 5G Base Station, WiFi Access Point, and LiFi Access Point. UEs, equipped with MPTCP, access services across 5G, WiFi, and LiFi networks simultaneously. We use 5G NR and WiFi 802.11ax modules, adhering to 3GPP Release-15 NR and IEEE 802.11 standards, while LiFi is modelled based on visible light communication. The Gauss-Markov mobility model simulates UE movement.

The UEs generate two types of application traffic with distinct packet sizes and data rate demands. We apply four legacy traffic steering algorithms: Priority-based active standby, Smallest Delay, Load Balancing, and Random Scheduling. Priority-based Active-Standby ranks links by priority, using the highest priority link unless unavailable. Smallest Delay uses the link with the lowest delay, and Load Balancing distributes traffic equally across available links. Experiments collect data from scenarios with 10, 20, and 30 UEs. The objective is to create an xAPP that classifies applications using a simple MLP with 25 inputs, three fully connected layers of 64 neurons, and two output classes: Application 1 and Application 2.

\subsubsection{Experiment Results}

\begin{figure*}
    \centering
    \begin{minipage}{0.48\linewidth}
\includegraphics[width=1\linewidth]{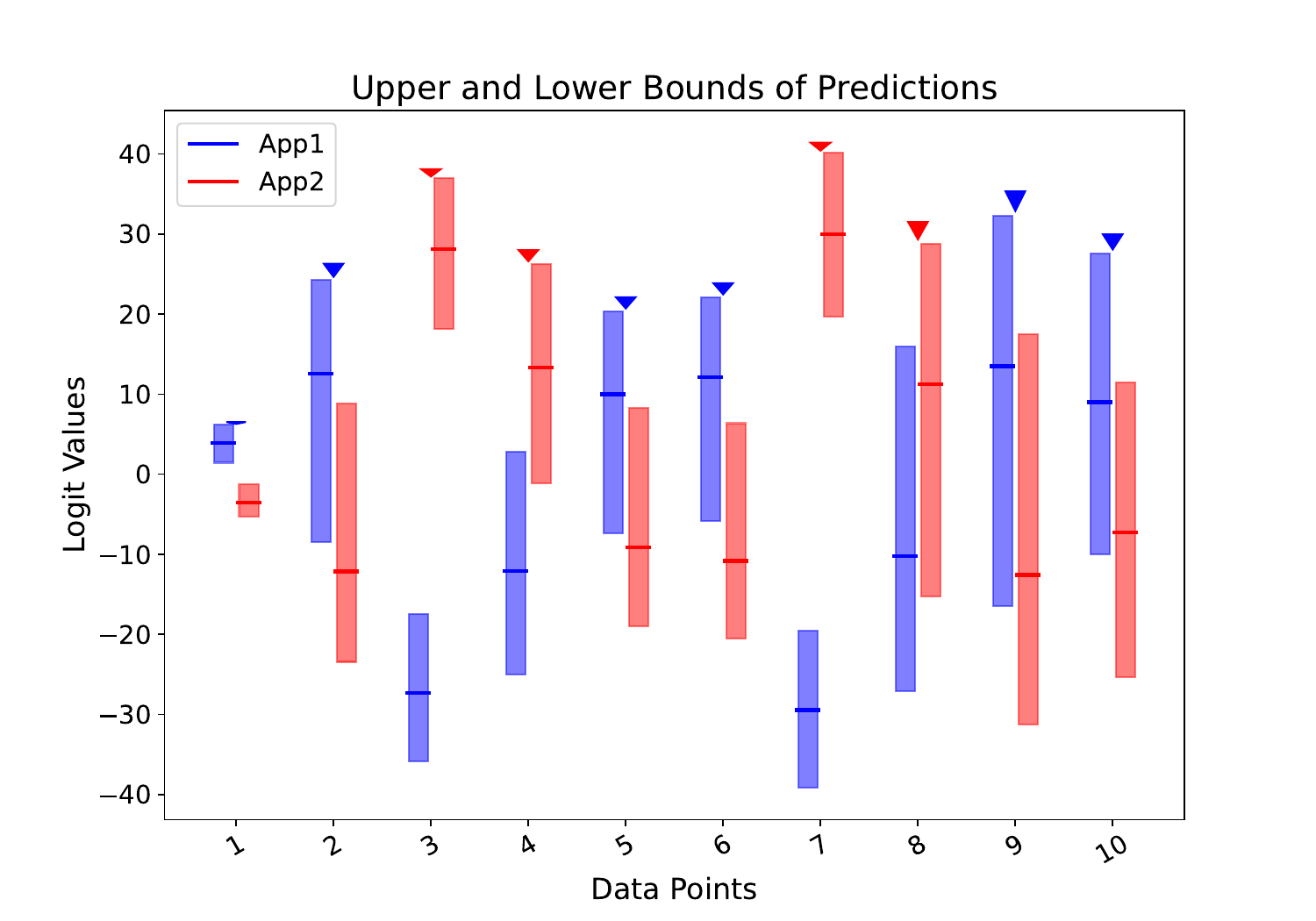}
        \centering
        (a) FVAI: Model Verification with AutoLiRPA
        \label{fig:fvai_exp}
    \end{minipage}
    \hfill
    \begin{minipage}{0.48\linewidth}
    \vspace{0.28cm}
        \includegraphics[width=1\linewidth]{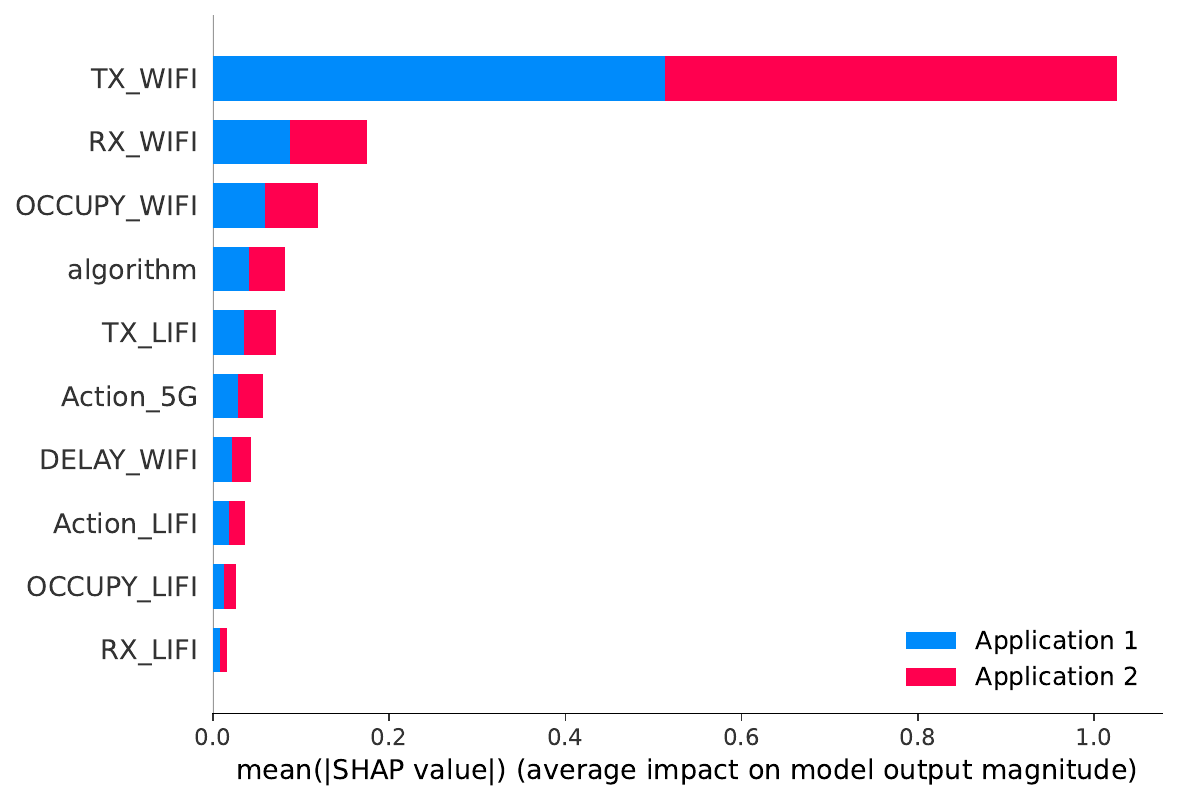}
         \centering
         (b) XAI: Model-Agnostic Explanation with SHAP
        \label{fig:XAI_FVAI}
    \end{minipage}
    \caption{Results of the Trustworthy AI modules}
    \label{fig:results}
\end{figure*}

The following subsection describes the trustworthy performance of the AI model within the DT MLOps pipeline, as illustrated in Figure \ref{fig:results}. 

\paragraph{FVAI}

Linear Relaxation based Perturbation Analysis (LiRPA) for neural networks, is a popular method that computes provable linear bounds of a neural network's output, given a certain amount of input perturbation. This has become a core component in robustness verification. CROWN \cite{xu:2020}, is a popular LiRPA algorithm that provides certified bounds on the output of neural networks under input perturbations.

We used the AutoLiRPA library’s CRWON implementation to verify the robustness of our MLP. Figure 5(a) depicts calculated bounds for ten sample data points in a two-class classification problem. The red and blue bars show the range of predicted logits when $L_{\infty}$ norm noise with a perturbation level of $0.05\%$ is added. The horizontal lines indicate the actual logit values without noise, while the coloured triangles at the top of each data point represent the true class labels. For data points 3 and 7, the logits are well separated with non-overlapping bounds, indicating good robustness against noise. Conversely, data points 2 and 4 show some overlap due to noise, reducing stability. Data points 8 and 9 exhibit significant overlap, suggesting reduced robustness.

We calculated the percentage of data points that have a separation between the bound bands, out of all the correctly predicted data, within the test dataset. With 95\% confidence identified, on average, even if there is a 0.05 perturbation to the input, the bounds bands of the predictions will not overlap for 50.13\% to 51.15\% of the correctly predicted data. This gave a good benchmark for the robustness of the model.

\paragraph{XAI}
SHAP (SHapley Additive exPlanations) is a method used to interpret the output of ML models by assigning each feature an importance value, known as a Shapley value. This value indicates the contribution of each feature to the prediction. In the experiment, SHAP is utilised to evaluate the significance of each network parameter in classifying user applications. Understanding feature importance helps to identify which parameters most influence the model's decisions, ensuring transparency and reliability. 

Figure 5(b) showcases, through Shapley values, that the most influential factor in the dataset is traffic transmitted via WiFi ($TX\_WiFi$), reflecting its key role in overall data rate variations driven by different application demands. The simulation scenario and traffic steering algorithm design favour WiFi for data transmission, making it a significant indicator of application types. Consequently, WiFi-related factors such as receiving data via WiFi ($RX\_WiFi$) and the occupied bandwidth of the WiFi link ($OCCUPY\_WiFi$) also play a crucial role in prediction accuracy. Additionally, the type of traffic steering algorithm ($algorithm$) affects traffic distribution across mATs. Since UEs consistently connect to 5G with lower priority, 5G-related metrics are less critical for classification. These insights are vital for optimising traffic steering algorithms and enhancing AI model reliability in 6G networks.

We obtained a classification accuracy of \textbf{99\%} using our MLP. Moreover, our approach supports both automated and manual validation based on outputs from the verification and explanation modules. The network operator can decide whether to publish the model as an xAPP if it meets the performance specifications defined by users. If the model meets these criteria, it is packaged into a container and deployed within the mATRIC. Once deployed, the xAPP functions as a real-time classifier for the mAT wireless network.

\section{Research Challenges and Open Issues}
\label{challenges}
In this section, we discuss the key outstanding challenges that require further investigation and outline way forward for potential future research directions.
\subsection{Challenges for Enabling Native Intelligence}
Implementing native intelligence in 6G requires a comprehensive, end-to-end approach and system-level considerations. AI performance in 6G can be affected by numerous factors, including different protocol stack layers, multiple network components, and diverse service demands. Effective lifecycle management of AI solutions is crucial for enabling native intelligence. MLOps is a promising tool, offering predefined pipelines that address lifecycle management needs. Although these pipelines are currently manually defined, automating their adaptation to evolving requirements remains challenging. Large Language Models (LLMs) could help by leveraging their global network knowledge to suggest relevant factors and the optimal sequence for pipeline inclusion.

\subsection{Challenges in AI Orchestration}
\subsubsection{Lack of a Structured AI/ML Task Description} In REASON, any AI/ML-related operation involving an AI model or a composition of models is described by employing an MLOps framework. As noted by Sculley~\emph{et al.}~\cite{sculley2015hidden}, the number of lines of code implementing a production-ready AI/ML model is rather small if compared to the sections of the code dealing with configuration issues, data collection, data verification, feature extraction from live inference requests, serving infrastructure and performance monitoring. One of the key tools to ensure all the components are codified in a non-ambiguous form is widely known as \emph{MLOps pipeline}.

From a graph-theoretical point of view, an MLOps pipeline is a directed graph structure consisting of several vertexes where the output of one vertex is the input of at most one district vertex. Regardless of the different embodiments and adaptations of the MLOps framework, the vertex set always consists of the following elements: i) Data Ingestion and Validation, ii) Model Train and Analysis, and iii)Model Execution and Monitoring. 
The MLOps framework currently lacks a system for managing the flow of a pipeline and allocating the necessary bare metal resources for running the pipeline's components. To address this issue, REASON introduces the concept of a \emph{pipeline task}. This can be defined as a valid path within the pipeline, potentially involving one or more pipeline components. Examples of pipeline tasks include Data Acquisition and Feature Engineering, as well as Model Performance Monitoring. Further investigation is needed to validate these concepts.

\subsubsection{Lack of the Resource Context Knowledge}

To enable effective end-to-end AI/ML orchestration, REASON requires each pipeline task to have associated execution requirements, including i) minimum GPU processing power, ii) minimum GPU memory, and iii) minimum memory capacity. Each task may also include a cost model detailing CPU/memory footprint and network usage over time. Additionally, tasks can be linked to monitoring metrics (e.g., maximum execution time, maximum prediction error) and monitoring patterns (e.g., average over time, periodic monitoring). Mapping pipeline tasks to suitable resources is equivalent to mapping microservices to computing nodes~\cite{Tassi} and requires further exploration.

\subsection{Challenges in Cognition and Trustworthy AI}
\subsubsection{Challenges of Cognition}

In COG, humans are expected to play a role, raising several issues: i) the extent to which humans should oversee AI decisions; ii) how to optimise human-machine interactions and task distribution across various autonomy levels; iii) how to adapt human roles as AI technologies evolve, including safely reducing human involvement in network operations. Additionally, the rise of LLMs presents opportunities for advanced functionalities within 6G networks. It is essential to consider how to maximise LLM capabilities in this context. While high-level COG requirements are outlined, implementing them system-wide and defining detailed interactions with AIO will require significant design and engineering efforts.

\subsubsection{Challenges of Trustworthy AI} 
It is vital to ensure the trustworthiness of the operation of a 6G network, which can be considered a complex system of systems. This complex system of systems, which will incorporate multiple AI supporting applications as well as non-AI functions all interacting with each other, will require multi-level analysis: at the component (i.e. a single AI model), subsystem, system and system-of-systems level. For example, multi-level verification and multi-level explainability might require the current state-of-the-art, which currently focuses on single AI models, to be advanced.  

Verification by testing has some limitations: i) it does not guarantee that a property (e.g., robustness to input perturbations) is satisfied by an AI model, unlike formal verification; ii) it can be computationally expensive; iii) generating test data for comprehensive coverage of the input domain can be challenging, and iv) achieving thorough test coverage of AI model internals is difficult. State-of-the-art formal verification methods also face limitations: i) they may not scale to large neural network models; ii) they may have constraints on the network architectures they can verify, and iii) they can be challenging to specify and prove required safety or trustworthiness properties within the formal framework.

AI trustworthiness will often require general compliance with laws and regulations that govern AI practices and data management. A flexible protocol for verifying this compliance is needed in order to adequately react to changes in AI rules and regulations that could occur over time.

\section{Conclusion}
\label{conclusion}
The transition to 6G networks requires advanced AI integration, ensuring high efficiency and trustworthiness. This paper presents the REASON project's framework, comprising AIO, COG, and AIM, for managing the AI lifecycle from development to deployment. This approach addresses the practical challenges of AI implementation in dynamic mAT environments while upholding trustworthy AI principles like reliability, accountability, and transparency. Continuous lifecycle management of AI models, including real-time monitoring and iterative updates via the DT, ensures their effectiveness and trustworthiness. The feasibility of this approach is demonstrated through xApps for content classification, optimising network operations while maintaining trust. Additionally, the paper discusses trustworthy AI principles, proposes an implementation framework for 6G, and identifies research challenges and open issues. Understanding and addressing these challenges are crucial for shaping the future of 6G networks and meeting evolving communication needs.

\section*{Acknowledgements}
This work is a contribution by Project REASON, a UK Government funded project under the Future Open Networks Research Challenge (FONRC) sponsored by the Department of Science Innovation and Technology (DSIT).



\newpage
\setcounter{figure}{0} %

\begin{figure*}
    \centering
    \includegraphics[width=0.85\textwidth]{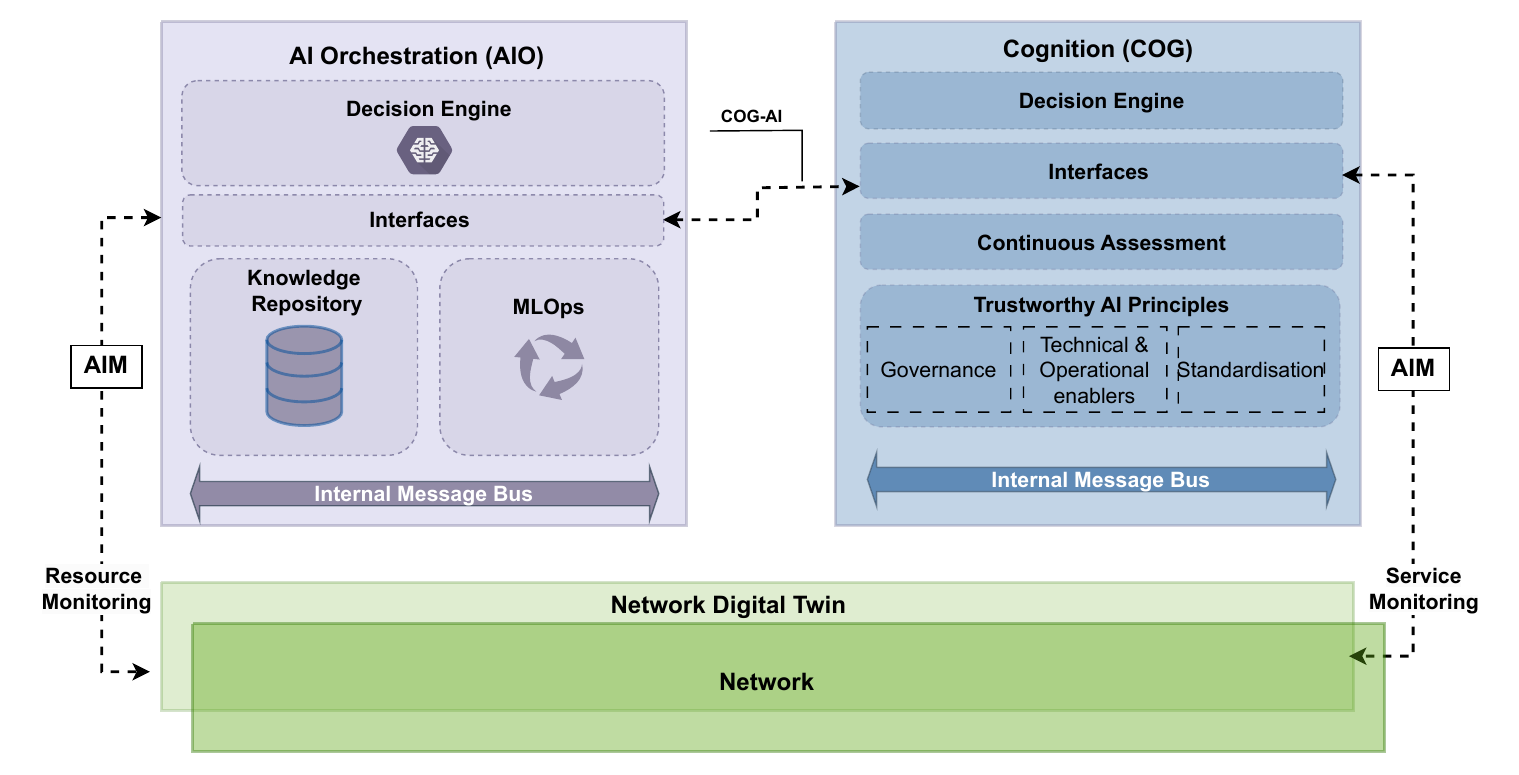}
    \caption{End-to-end AI Plane: AIO, COG, AIM and DT}
\end{figure*} 

\newpage

\begin{figure*}
    \centering
    \includegraphics[width=0.85\linewidth]{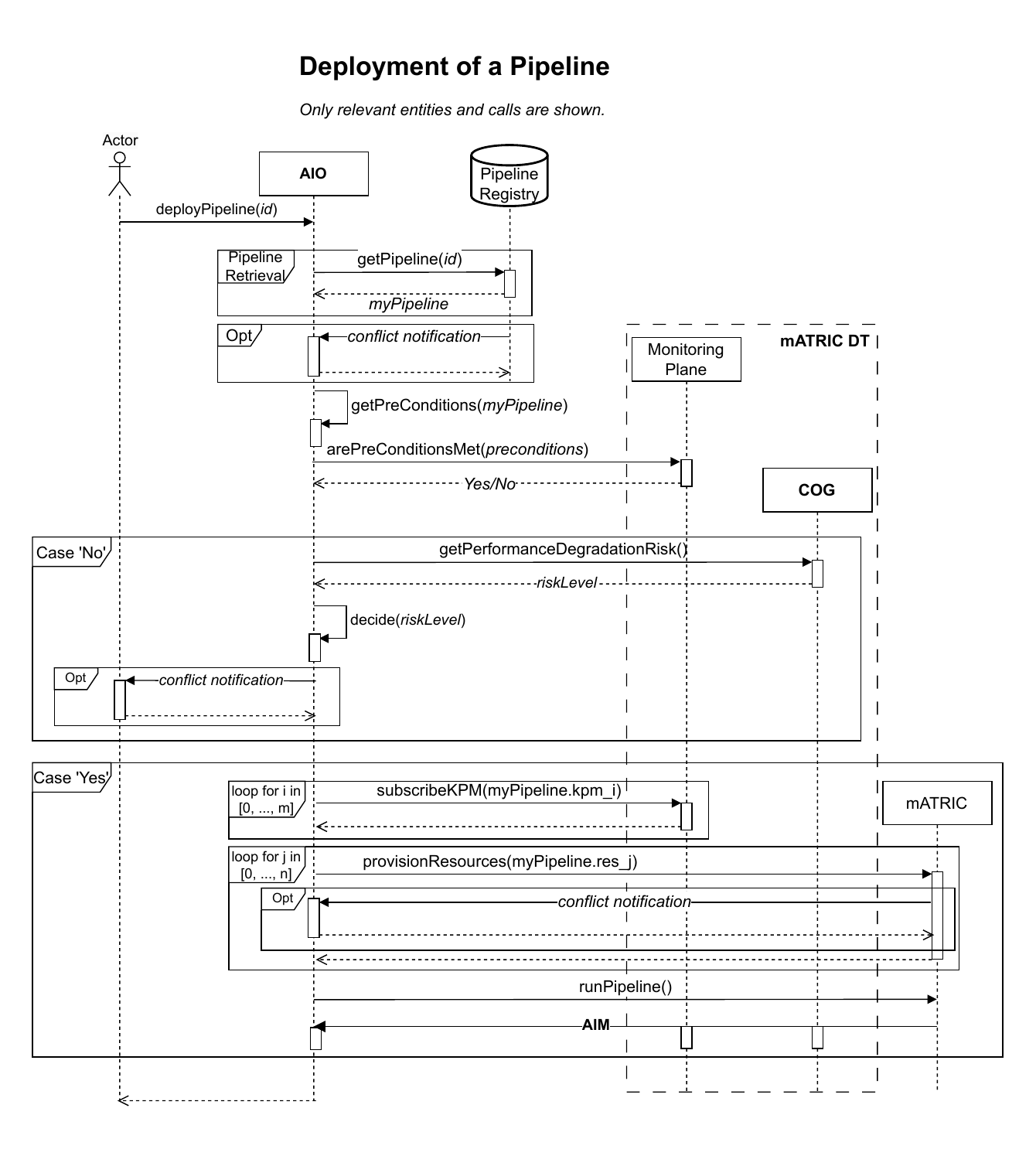}
    \caption{Example Deployment of an AI Pipeline using AIO, COG, AIM and the mATRIC DT}
    \vspace{-0.2cm}
\end{figure*}

\newpage

\begin{figure*}[!t]
    \centering
\includegraphics[width=0.95\linewidth]{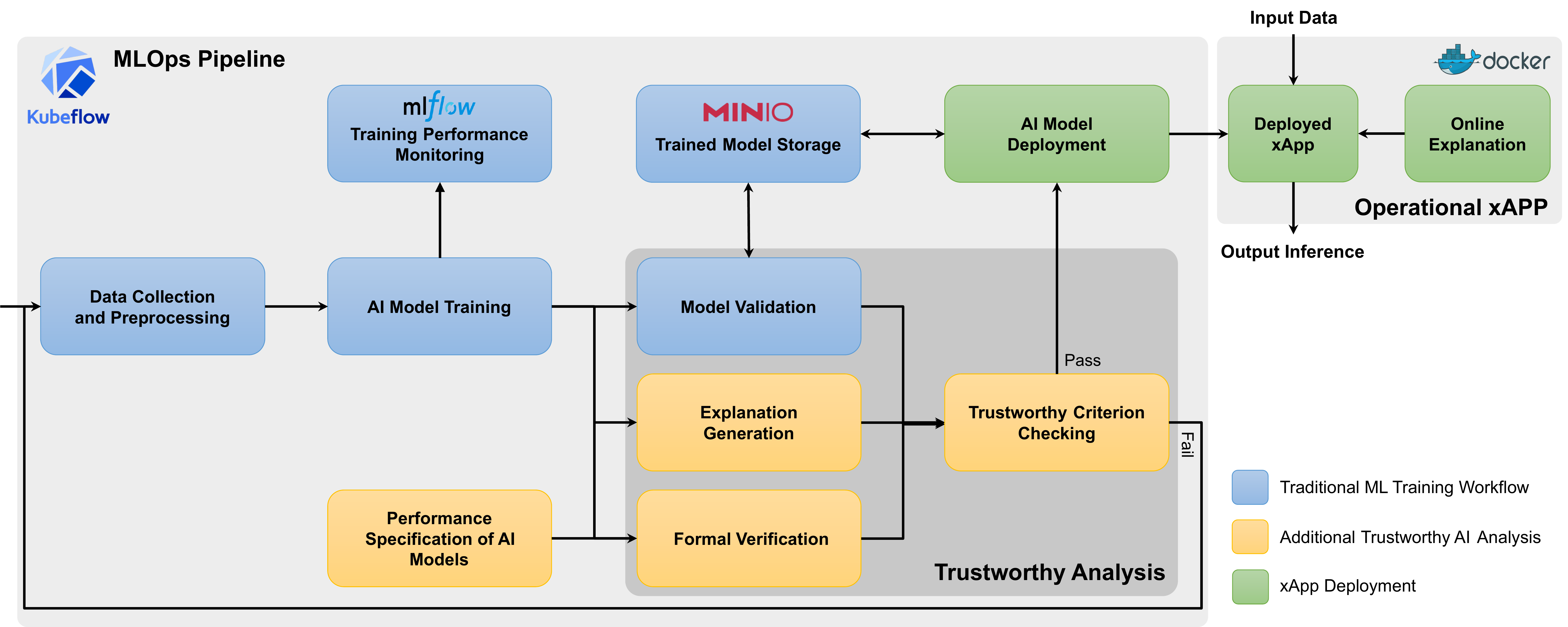} 
    \caption{Testbed Deployment of an MLOps Pipeline with Trustworthy Evaluation}
\end{figure*}

\newpage

\begin{figure}[!t]
    \centering
    \includegraphics[width=0.85\linewidth]{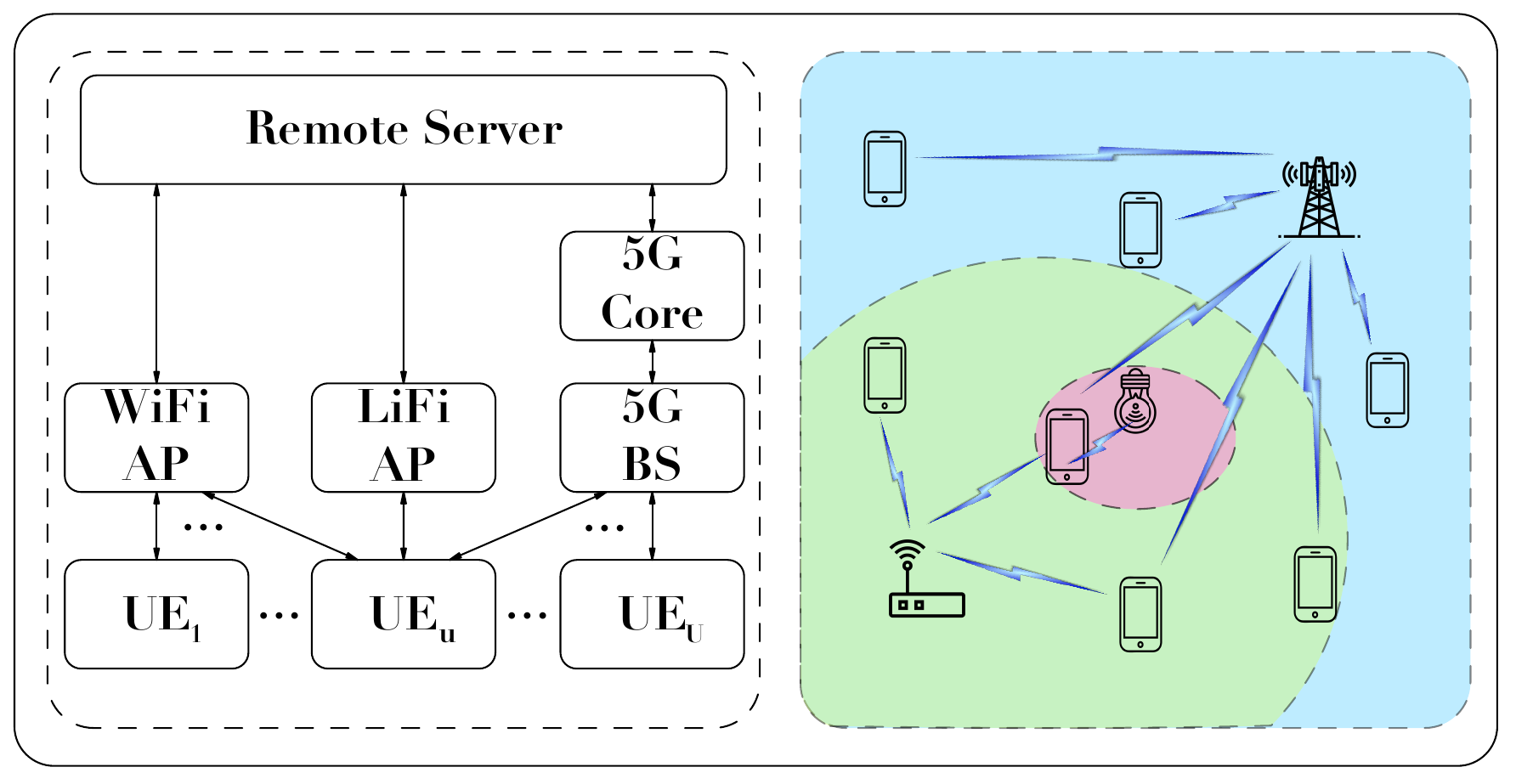}
    \caption{mAT Experiment in NS3 Simulator}
\end{figure}

\newpage

\begin{figure*}
    \centering
    \begin{minipage}{0.48\linewidth}
\includegraphics[width=1\linewidth]{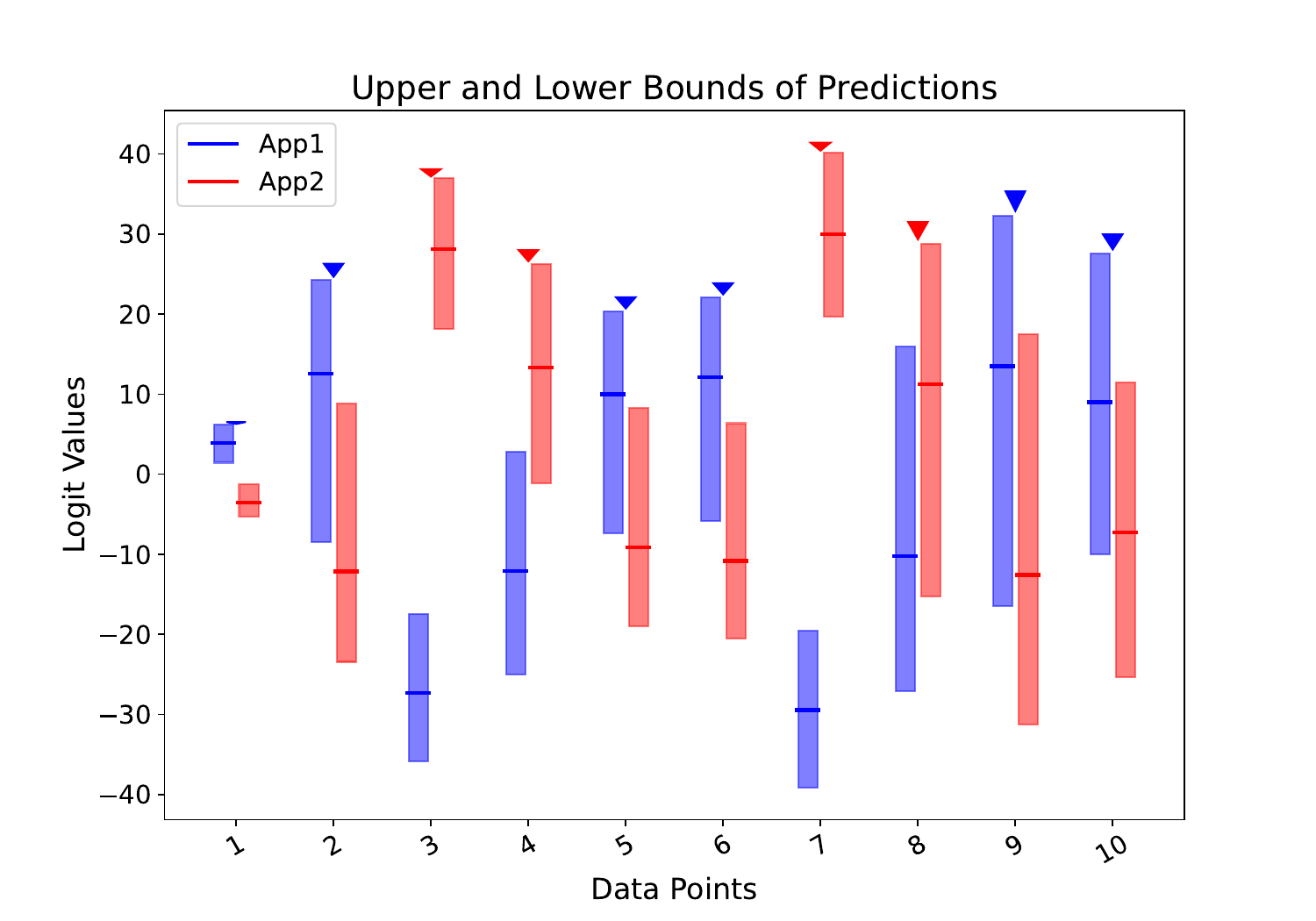}
        \centering
        (a) FVAI: Model Verification with AutoLiRPA
    \end{minipage}
    \hfill
    \begin{minipage}{0.48\linewidth}
    \vspace{0.28cm}
        \includegraphics[width=1\linewidth]{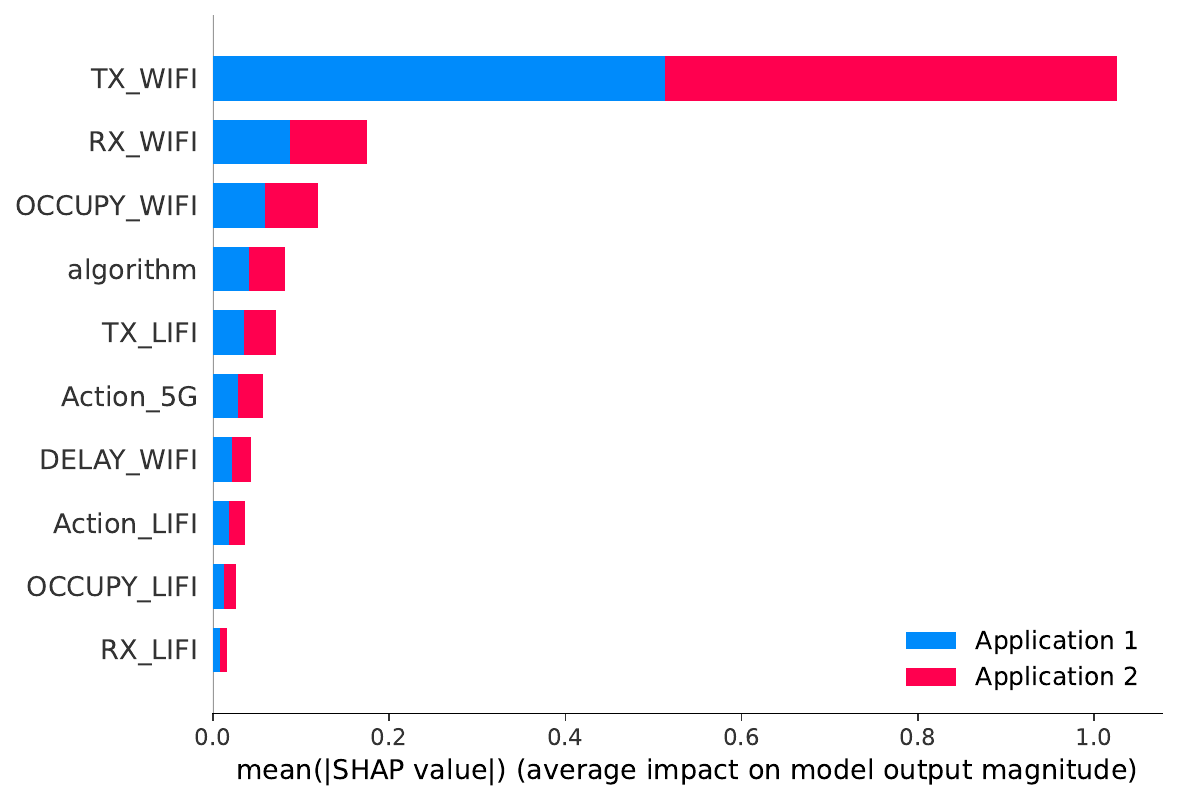}
         \centering
         (b) XAI: Model-Agnostic Explanation with SHAP
    \end{minipage}
    \caption{Results of the Trustworthy AI modules}
\end{figure*}


\end{document}